\documentclass[twocolumn,preprintnumbers,amsmath,amssymb]{revtex4-2}

\usepackage{graphicx}
\usepackage{amsmath}
\usepackage{algorithm}
\usepackage{algorithmic}
\newcommand{\commentoutA}[1]{}

\begin{document}

\preprint{LA-UR-24-29911}



\title{Susceptibility Formulation of Density Matrix Perturbation Theory}






\author{Anders M.\ N. Niklasson}
\email{amn@lanl.gov}
\author {Adela Habib}
\author{Joshua D. Finkelstein}
\affiliation{Theoretical Division, Los Alamos National Laboratory, Los Alamos, New Mexico 87545, USA}
\author{Emanuel H. Rubensson}

\affiliation{Division of Scientific Computing, Department of Information Technology, Uppsala University, Box 337, SE-751 05 Uppsala, Sweden}

\date{\today}

\begin{abstract}
Density matrix perturbation theory based on recursive Fermi-operator expansions provides a computationally efficient framework for time-independent response calculations in quantum chemistry and materials science.
From a perturbation in the Hamiltonian we can calculate the first-order perturbation in the density matrix, which then gives us the linear response in the expectation values for some chosen set of observables. Here we present an alternative, {\it dual} formulation, where we instead calculate the static susceptibility of an observable, which then gives us the linear response in the expectation values for any number of different Hamiltonian perturbations. We show how the calculation of the susceptibility can be performed with the same expansion schemes used in recursive density matrix perturbation theory, including generalizations to fractional occupation numbers and self-consistent linear response calculations, i.e. similar to density functional perturbation theory. As with recursive density matrix perturbation theory, the dual susceptibility formulation is well suited for numerically thresholded sparse matrix algebra, which has linear scaling complexity for sufficiently large sparse systems. Similarly, the recursive computation of the susceptibility also seamlessly integrates with the computational framework of deep neural networks used in artificial intelligence (AI) applications. This integration enables the calculation of quantum response properties that can leverage cutting-edge AI-hardware, such as Nvidia Tensor cores or Google Tensor Processing Units. We demonstrate performance for recursive susceptibility calculations using Nvidia Graphics Processing Units and Tensor cores.
\end{abstract}

\maketitle

\section{Introduction}

Quantum perturbation theory provides a framework to calculate response properties of chemical systems. These calculations can be performed, for example, using methods such as Rayleigh-Schr\"{o}dinger perturbation theory, coupled perturbed self-consistent field methods, or density functional perturbation theory \cite{THelgaker02,Stevens63, Hirchfelder64,Gerratt68,JPople79,HSekino86,SKarna91,SBaroni01}.

An alternative approach to calculating response properties in quantum chemistry is to use density matrix-based perturbation theory, which dates back to the pioneering work of McWeeny in the 1960s for time-independent response properties \cite{RMcWeeny62}, and to the work of scientists such as Adler and Wiser for calculating the frequency-dependent dielectric constant in solids based on the time-dependent quantum Liouville equation \cite{Adler62,Wiser63}.

In time-independent density matrix perturbation theory based on recursive Fermi-operator expansions \cite{ANiklasson04,VWeber04,VWeber05,Niklasson2011,ANiklasson15,Truflandier20,Shang21,H_Shang21,JFinkelstein22,Chen22}, linear response properties are determined from the first-order perturbation in the single-particle density matrix induced by a time-independent perturbation to the Hamiltonian. To compute the first-order response in the density matrix, a first-order perturbation is calculated at each step of the recursive Fermi-operator expansion \cite{ANiklasson04}. The resulting first-order response in the density matrix can then be used to evaluate the linear response of expectation values for a wide range of observables in response to the given Hamiltonian perturbation.

Other time-independent density matrix perturbation approaches have also been developed, for example, where the linear response density matrices appear as solutions to Sylvester-like commutator (or super-operator) equations \cite{COchsenfeld04,Liang05,JKussmann07,Ringholm14,Kussman15}. However, these methods are not considered in this article.

In this article, we present an alternative {\it dual} approach to time-independent density matrix perturbation theory based on recursive Fermi-operator expansions. In this dual formulation, the linear response of an observable's expectation value is determined from the static susceptibility of the observable. Once the susceptibility is calculated, the response in the expectation value can be readily obtained for any number of different perturbations to the underlying Hamiltonian. We will demonstrate how this dual approach can employ the same recursive scheme to compute observable susceptibilities that has previously been used to calculate the density matrix response in recursive density matrix perturbation theory. This susceptibility-based linear response theory provides an efficient and useful alternative that complements previous density matrix approaches, particularly when multiple response calculations are needed for the same observable under different Hamiltonian perturbations. 

The computational structure of the recursive density matrix perturbation and susceptibility calculations aligns seamlessly with that of deep neural networks commonly used in artificial intelligence (AI) applications \cite{JFinkelstein21,JFinkelstein22}. 
As a result, these density matrix response calculations can be performed with high efficiency by leveraging Graphics Processing Units (GPUs) as well as state-of-the-art AI hardware accelerators such as Tensor Processing Units or Tensor cores \cite{MCawkwell12b,ANiklasson16,ALewis22,RPederson22,JFinkelstein21,JFinkelstein21B,RSchade22,JFinkelstein22,AHabib24}. 
Additionally, density matrix perturbation theory is well-suited for computations using numerically thresholded sparse matrix algebra \cite{ANiklasson04,VWeber04,VWeber05,COchsenfeld04, Shang21,H_Shang21,Xin22}, enabling linear-scaling computational complexity for sufficiently large non-metallic systems \cite{Goedecker99,DBowler12}. This is facilitated by a computational kernel based on sparse matrix-matrix multiplications
\cite{FGustavson78,ERubensson05,ERubensson08a,BML,ABuluc12,NBock13,SMniszewski15}.
The dual susceptibility formulation of density matrix perturbation theory presented in this article shares the same computational structure and should therefore also be capable of achieving linear-scaling complexity.

The recursive density matrix perturbation theory and its dual susceptibility formulation are suitable for effective single-particle methods such as Hartree-Fock, density functional theory (DFT) \cite{CRoothaan51,hohen,KohnSham65}, or approximate DFT \cite{MElstner98,MFinnis98,BHourahine20} as well as semi-empirical methods \cite{MDewar77,MDewar85,JStewart13,CBannwarth18,PDral19,WMalone20,ZGuoqing20,CBannwarth20}. 
Our presentation of the susceptibility formulation of density matrix perturbation theory is first focused on {\it non-self-consistent} linear response calculations at zero electronic temperature with integer occupations of the electronic states. However, our approach is general and applicable also to coupled perturbed self-consistent field or density functional perturbation theory \cite{Gerratt68,JPople79,SBaroni01}, as well as for systems at elevated electronic temperatures with fractional occupation numbers of the electronic states. The presentation of the theory is mainly focused on the recursive Fermi-operator expansions using second-order spectral projections (SP2) \cite{ANiklasson02,EHRubensson11,EHRubensson14,SMMniszewski19,JFinkelstein22} and the corresponding density matrix perturbation theory \cite{ANiklasson04,VWeber04,JFinkelstein22}. The SP2 scheme is a particularly computationally and memory efficient approach, especially for calculations using GPUs and Tensor cores \cite{MCawkwell12b,JFinkelstein21}. However, most alternative recursive Fermi-operator schemes should also be applicable, including expansions for fractional occupation numbers \cite{APalser98,KNemeth00,DMazziotti03,ANiklasson03,ANiklasson03B,ANiklasson2011,ANiklasson15,Truflandier16,JFinkelstein21,JFinkelstein23}. 

In our discussion we assume the operators and observables are represented with a finite local real-valued basis-set representation as $N \times N$ matrices ($ \in \mathbb{R}^{N \times N}$). In this way all algorithms in our presentation involve straightforward matrix operations. If not otherwise stated, we assume an orthonormal basis-set, where the basis-set $N \times N$ overlap matrix, $S$, is equal to the identity matrix, $I$, i.e.\ $S = I$. In general, this also means that operators, observables, overlap and density matrices, are all symmetric. 

After presenting the SP2 recursive Fermi-operator (or step function) expansion scheme for the effective single-particle density matrix and its first-order response at zero electronic temperature, we present the dual susceptibility formulation using the very same recursive framework. We then discuss generalizations to non-orthonormal representations with position-dependent response properties and finite temperature susceptibilities for systems with fractional occupation numbers, as well as application to self-consistent field (SCF) theory. We also demonstrate the numerical equivalence between the two dual approaches in our recursive linear response theory. Our dual susceptibility formulation of the recursive density matrix perturbation theory is then formulated with mixed precision using low precision floating-point operations. At the end, we demonstrate performance of the recursive susceptibility calculations using GPUs and Tensor cores. The Appendix contains some complementary derivations and algorithms.

\section{Recursive Fermi-operator expansion}

 In effective single-particle electronic structure theory, the  density matrix, $D^{(0)} \in {\mathbb R}^{N \times N}$, for a given symmetric (or Hermitian) Hamiltonian matrix, $H^{(0)}  \in {\mathbb R}^{N \times N}$, (using some suitable basis-set representation) is given by
 \begin{equation} 
     D^{(0)} = \theta\left(\mu I - H^{(0)}\right)\;,
 \end{equation}
 where $\theta(X): {\mathbb R}^{N\times N}\rightarrow {\mathbb R}^{N \times N}$ is the Heaviside matrix step function, representing the Fermi function at zero electronic temperatures. The matrix $I$ is the identity matrix and $\mu$ is the chemical potential, chosen such that the trace of the density matrix is the number of occupied states, $N_{\rm occ}$, i.e.\ ${\rm Tr}[D^{(0)}] = N_{\rm occ}$. In a recursive Fermi-operator expansion based on second-order spectral projections, i.e.\ the SP2 scheme \cite{ANiklasson02,EHRubensson11,EHRubensson14,SMMniszewski19,JFinkelstein21}, the density matrix, $D^{(0)}$, is given by 
 \begin{align} \label{DM0}
     &X_1 = \alpha I + \beta H^{(0)}  \nonumber \\
     &{\rm for~} n = 1 {\rm ~to~} M\nonumber \\
     & ~~~~ X_{n+1} = (1-\sigma_n)X_n + \sigma_nX_n^2\\
     &{\rm end}\nonumber \\
     &D^{(0)} = X_{M+1}.\nonumber 
 \end{align} 
The initial linear transformation of $H^{(0)}$ to $X_1$, which is defined by the scalar values of $\alpha$ and $\beta$, is chosen so that the eigenvalue spectrum of $H^{(0)}$ is transformed in reversed order to the interval $[0,1]$.  The particular values for $\alpha$ and $\beta$ are then given by $\alpha = \varepsilon_{\rm max}(\varepsilon_{\rm max}-\varepsilon_{\rm min})^{-1}$ and $\beta =-(\varepsilon_{\rm max}-\varepsilon_{\rm min})^{-1}$, where $\varepsilon_{\max}$ and $\varepsilon_{\min}$ are estimates of the upper and lower bounds for the eigenvalue spectrum of $H^{(0)}$, respectively. The scalar factors, $\sigma_n = \pm 1$, are selected in each step so that the trace of $X_{n+1}$ is as close as possible to $N_{\rm occ}$  after each new iteration. The value of the number of recursion steps, $M$, is chosen to be sufficiently high such that $D^{(0)}$ is idempotent with all eigenvalues 1 or 0 (within some numerical tolerance). This convergence can be determined by parameter-free conditions that work even in the case of low precision floating point arithmetic and approximate numerically thresholded sparse matrix algebra \cite{AKruchinina16,JFinkelstein21}. 

From the converged density matrix, $D^{(0)}$, in Eq.\ (\ref{DM0}), the expectation value, $a^{(0)}$ of an observable $A$ can be determined from 
 \begin{equation}
     a^{(0)} = \langle A \rangle \equiv {\rm Tr}\left[ A D^{(0)}\right]. \label{expectation}
 \end{equation}
The observable, $A$, can be, for example, the matrix corresponding to the dipole moment operator, the approximate electron occupation of an atom, or various energy terms. In this way a number of different expectation values can be determined directly from the density matrix.

The recursive density matrix expansion in Eq.\ (\ref{DM0}) is based on repeated generalized matrix-matrix multiplications. These multiplications can be performed using numerically thresholded sparse matrix algebra to achieve reduced linear scaling complexity for sufficiently large sparse matrices \cite{SGoedecker99,DBowler12}. The generalized matrix multiplications performed in each iteration of the SP2 algorithm correspond to the tensor contractions typical of a deep neural network and can be performed with high-performance on GPUs or specialized AI hardware using dense matrix algebra \cite{MCawkwell12b,JFinkelstein21,ALewis22,RPederson22}. These same computational advantages hold for the density matrix perturbation and susceptibility calculations discussed below. 

For sufficiently large sparse problems a sparse linear scaling approach will always be faster than using cubically scaling dense matrix algebra. However, for intermediate sized problems, including up to a few thousand atoms, dense matrix algebra calculations using GPUs or AI hardware is often the better choice \cite{JFinkelstein21,AHabib24}. Nevertheless, if we use, for example, divide and conquer \cite{WYang91,WYang95,WPan98,OWarschkow98,KKitaura99,FShimojo08,MKobayashi11,YNishimoto2014} or graph-based linear-scaling electronic structure theory \cite{ANiklasson16,Djidjev16,MLAss18,Djidjev19,MLass20,RSchade22,CNegre22}, dense matrix operations can be used also for very large systems including hundred's of thousands of atoms. Graph-based linear scaling electronic structure theory is based on a one-to-one mapping between a numerically thresholded sparse matrix algebra and a divide and conquer approach. In this case the graph of the electronic entanglement of a molecular system is partitioned into smaller partially overlapping subgraphs, where the density matrix of each partitioning can be calculated with dense matrix algebra using the recursive Fermi-operator expansion scheme running on GPU's or AI hardware \cite{MCawkwell12b,ANiklasson16}. The subgraphs may include up to a few thousand atoms, depending on the choice of the graph partitioning and the basis set.

\section{Density Matrix Perturbation Theory}

Density matrix perturbation theory based on recursive Fermi-operator expansion schemes \cite{ANiklasson04,VWeber04} allows us to calculate response properties from effective-single particle electronic structure theory, such as DFT or Hartree-Fock based methods. In general these methods require an iterative self-consistent field solution \cite{Gerratt68,JPople79,SBaroni01,VWeber04,COchsenfeld04,VWeber05,Liang05,JKussmann07,Ringholm14}. We will first focus on non-self-consistent response calculations using orthonormal basis-set representations with integer occupation of the electronic states. Generalizations to self-consistent calculations and elevated electronic temperatures with fractional occupation numbers and non-orthonormal basis-set representations are straightforward and will thereafter be demonstrated in section \ref{Gen} \cite{VWeber05,ANiklasson15}.

If we introduce a time-independent, first-order perturbation to the Hamiltonian, $H^{(1)} \in \mathbb{R}^{N\times N}$, where
\begin{equation}
    H = H^{(0)} + \lambda H^{(1)}\;,
\end{equation}
we can calculate the corresponding time-independent linear response in the density matrix, $D^{(1)}$, as a Gateaux directional derivative in the direction of $H^{(1)}$, 
\begin{align}
D^{(1)} &= \frac{d \theta\left(\mu I - (H^{(0)} + \lambda H^{(1)})\right)}{d \lambda}\Big \vert_{\lambda = 0},   
\end{align} 
using a recursive expansion based on Eq.\ (\ref{DM0}). This calculation can then be performed with recursive density matrix perturbation theory \cite{ANiklasson04} using the algorithm,
 \begin{align}\label{DM1}
     &Y_1 = \beta H^{(1)}\nonumber \\
     &{\rm for~} n = 1 {\rm ~to~} M\nonumber \\
     & ~~~~ Y_{n+1} = (1-\sigma_n)Y_n + \sigma_n\left(Y_nX_n + X_nY_n\right)\\
     &{\rm end}\nonumber \\
     &D^{(1)} = Y_{M+1}\nonumber\;.
 \end{align}
Here we use the same values of $\beta$, $M$, $\{X_n\}$, and $\{\sigma_n\}$ ($n=1,2,\ldots ,M$) as for the ground state calculation of $D^{(0)}$ in Eq.\ (\ref{DM0}). It is easy to combine Eq.\ (\ref{DM1}) and Eq.\ (\ref{DM0}) to avoid having to store the matrices $\{X_n\}$ (See Eq.\ (\ref{DMPert}) in the Appendix). 

For any set of observables $\{A\}$, the response in the expectation value from the perturbation, $H^{(1)}$, is then given by
\begin{equation}\label{direct}
     a^{(1)} = \frac{d \langle A \rangle}{d \lambda} 
     \Big \vert_{\lambda = 0} = {\rm Tr} \left[ AD^{(1)}\right].
\end{equation} 
Given the linear Hamiltonian perturbation, $H^{(1)}$, we can thus calculate the density matrix perturbation, $D^{(1)}$, in Eq.\ (\ref{DM1}), from which we then can determine the response in any chosen set of single-particle observables, $\{A\}$, in Eq.\ (\ref{direct}). However, for each new Hamiltonian perturbation, $H^{(1)}$, it is necessary to calculate a completely new response in the density matrix, $D^{(1)}$. With the dual susceptibility formulation, presented below, this can be avoided.

\section{Susceptibility formulation of density matrix perturbation Theory}

Recalculating the response in the density matrix, $D^{(1)}$, for each new perturbation in the Hamiltonian, $H^{(1)}$, is a very inefficient approach if we need multiple response calculations with different perturbations, $H^{(1)}$, for the same observable, $A$. As an alternative to the response calculation in Eq.\ (\ref{direct}), we therefore propose a complementary dual approach, where we instead calculate the linear response of a given single observable, $A$, for any chosen set of different Hamiltonian perturbations. 

To understand how this is possible we can look at the change in the observable, $a = a^{(0)} + \delta a$, with respect to some small variation in the Hamiltonian, $H = H^{(0)} + \delta H$. For small changes in $H$ this can be described by a linearization around $H^{(0)}$, where
\begin{equation}
     a \approx a^{(0)} + \delta a = {\rm Tr} \left[A D^{(0)} \right] + {\rm Tr}\left[ \delta H\frac{\partial {\rm Tr}[AD^{(0)}]}{\partial H} \right].
\end{equation}
If we define the {\it susceptibility}, $\chi^A$, for an observable, $A$, by
\begin{equation}
 \chi^A_{ij} \equiv \frac{ \partial \langle A \rangle}{\partial H_{ij}}  = \frac{\partial {\rm Tr}[AD^{(0)}]}{\partial H_{ij}},\label{Suscept}
\end{equation}
we then find that the first-order response, $a^{(1)}$, in the observable from a first-order perturbation in the Hamiltonian, $H = H^{(0)} + \lambda H^{(1)}$, is given by
\begin{equation} \label{dual}
     a^{(1)} = \frac{d \langle A \rangle}{d \lambda} 
     \Big \vert_{\lambda = 0} = {\rm Tr} \left[\chi^AH^{(1)}\right] \;.
\end{equation}
The first-order response of the observable, i.e.,
\begin{equation}
    a(\lambda) = a^{(0)}+ \lambda a^{(1)},
\end{equation}
is thus given through the linear susceptibility, $\chi^A$, in the single-particle expectation value for observable $A$, and is valid for any chosen first-order response in the Hamiltonian, $H^{(1)}$. In this way, we avoid the density matrix perturbation calculation for each new perturbation in the Hamiltonian, $H^{(1)}$. Instead we only need a single susceptibility calculation of the observable, $A$, for each new perturbation in the Hamiltonian. Furthermore, in case $A$ is more localized than $H^{(1)}$, the calculation of $\chi^A$ may be much cheaper than the calculation of $D^{(1)}$, if sparse matrix algebra is used \cite{ANiklasson04}. Therefore, the susceptibility calculation sometimes may be preferred also for the calculation of a single expectation value. 
\begin{figure}
\includegraphics[scale=0.25]{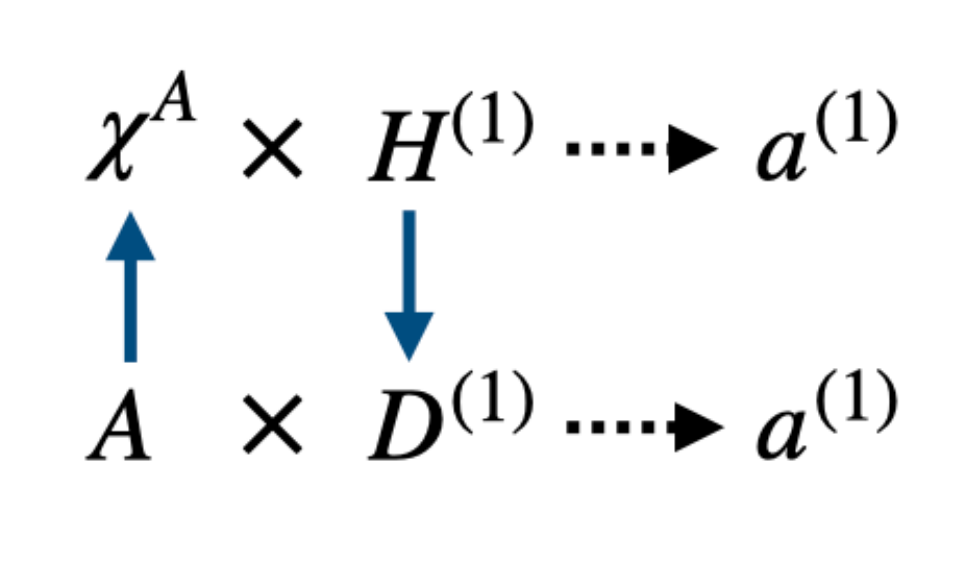}
\caption{\label{Fig_1}
{\small The duality relations in the calculation of the linear response of an expectation value, $a^{(1)}$, either using Eq.\ (\ref{direct}) or Eq.\ (\ref{dual}). The vertical solid vectors denote recursive density matrix perturbation expansions, using either $A$ or $H^{(1)}$ as first-order perturbations, and $\chi^A$ or $D^{(1)}$ as their linear responses. The density matrix perturbation approach in the lower line is suitable if we need to know the linear response for a number of different observables, $A$, whereas the upper line represents the dual susceptibility formulation, which is well adapted if we need to calculate the linear response for a number of different perturbations, $H^{(1)}$, or, for example, if $A$ is a more localized (single-atom) property than $H^{(1)}$. }}
\end{figure}

The dual picture for the susceptibility-based calculation is illustrated in Fig.\ \ref{Fig_1}. The main objective of this article is to demonstrate how we can perform response calculations with the dual susceptibility-based approach illustrated in Fig.\ \ref{Fig_1} using the same kind of recursive density matrix perturbation expansions as in Eq.\ (\ref{DM1}). 

It may appear that a straightforward calculation of the susceptibility matrix elements, $\{\chi_{ij}^A\}$, in Eq.\ (\ref{Suscept}), requires a response calculation for each component in the Hamiltonian, $\{H_{ij}\}$, but this is not necessary. Instead, we can calculate the full susceptibility matrix, $\chi^A$, directly. We will provide two separate, but mathematically equivalent ways to calculate the susceptibility -- using either a backward or a forward susceptibility expansion scheme.

 \subsection{Backward susceptibility expansion}
 
 To calculate the full susceptibility matrix, $\chi^A$, for the observable $A$ (we here assume $A= A^T$), we can use the same recursion as for the density matrix response calculations in Eq.\ (\ref{DM1}), but in the reverse backward direction, starting with $Y_M = A$ and $n = M$, i.e.\ where
  \begin{align} \label{DM2}
     &Y_M = A \nonumber\\
     &{\rm for~} n = M {\rm ~to~} 1\nonumber \\
     & ~~~~ Y_{n-1} = (1-\sigma_n)Y_n + \sigma_n(Y_nX_n + X_nY_n)\\
     &{\rm end}\nonumber \\
     &\chi^A = \beta Y_0.\nonumber
 \end{align}
The same values of $\beta$, $M$, $\{X_n\}$, and $\{\sigma_n\}$ are used as for the ground state calculation of $D^{(0)}$ in Eq.\ (\ref{DM0}). The algorithm in Eq.\ (\ref{DM2}) is the recursive differentiation of the observable, $a^{(0)} = \langle A\rangle$, in Eq.\ (\ref{expectation}) with respect to the changes in the Hamiltonian matrix elements (as in Eq.\ (\ref{Suscept})), using the chain-rule for derivatives for the expansion in Eq.\ (\ref{DM0}). This methodology is currently popular in machine learning and the corresponding techniques are then typically referred to as ``backpropagation'' or automatic differentiation, where the differentiation often is performed automatically through software support.

The recursive susceptibility expansion in Eq.\ (\ref{DM2}) is well suited for efficient calculations, either using numerically thresholded sparse matrix algebra or dense matrix operations with GPUs or AI-hardware accelerators \cite{ANiklasson04,VWeber04,VWeber05,JFinkelstein22}.

In contrast to the density matrix perturbation in Eq.\ (\ref{DM1}), where the sequence $\left\{ X_n \right\}_{n=1}^M$ can be generated on-the-fly by merging Eq.\ (\ref{DM0}) and Eq.\ (\ref{DM1}), we need to store the $\left\{ X_n \right\}_{n=1}^M$ sequence in advance for the recursive susceptibility calculation in Eq.\ (\ref{DM2}). This storage is particularly limiting if we apply the approach to self-consistent calculations or for applications using specialized AI-hardware where we are limited by the size of the accessible memory. The additional storage can be avoided if we can replace the backward expansion using a forward approach.

\subsection{Forward susceptibility expansion}

To construct a forward recursive expansion of the susceptibility we will use the relation (see Appendix):
 \begin{equation}
\frac{\partial {\rm Tr}\left[Af(X)\right]  }{\partial X} =  
\frac{d f(X + \lambda A^T)}{d \lambda}\Big \vert_{\lambda = 0}, \label{MatFuncRel}
 \end{equation}
 which holds for symmetric matrices, $X = X^T$. 
 In this relation we can identify the matrix $A$ with the observable, which we may also assume is symmetric, and replace the matrix function $f(X)$ by the density matrix $D^{(0)}$, which is generated by
 the Heaviside step function of $X = \mu I - H^{(0)}$ (or the Fermi function of $X =  H^{(0)}$). In this way we find from Eq.\ (\ref{Suscept}) that the susceptibility is given by 
 \begin{equation}
 \chi^A_{ij} \equiv \frac{\partial \langle A \rangle }{\partial H_{ij}} =  \left\{\frac{d \theta(\mu I - H - \lambda A^T)}{d \lambda}\Big \vert_{\lambda = 0}\right\}_{ij}. \label{Suscept_Der}
 \end{equation}
This means that the susceptibility calculation in Eq.\ (\ref{Suscept}) or (\ref{DM2}) is equivalent to a density matrix response calculation as in Eq.\ (\ref{DM1}), where the Hamiltonian perturbation, $H^{(1)}$, is replaced by the observable, $A$. This also means that
\begin{align}\label{Equivalence}
   a^{(1)} =  {\rm Tr} \left[A D^{(1)}\right] &= {\rm Tr} \left[\chi^A H^{(1)}\right],
\end{align}
where the susceptibility matrix, $\chi^A$, of the observable, $A = A^T$, can be generated by
  \begin{align} \label{DM3}
     &Y_1 = \beta A \nonumber\\
     &{\rm for~} n = 1 {\rm ~to~} M\nonumber \\
     & ~~~~ Y_{n+1} = (1-\sigma_n)Y_n + \sigma_n(Y_nX_n + X_nY_n)\\
     &{\rm end}\nonumber \\
     &\chi^A = Y_{M+1}.\nonumber
 \end{align}
The advantage with this forward expansion is that the sequence $\{X_n\}_{n=1}^M$ can be calculated on-the-fly through Eq.\ (\ref{DM0}), without the need of any temporary storage (See Appendix). This can be of particular importance in high-performance calculations using GPUs and specialized accelerated AI hardware, such as tensor cores -- especially for larger systems. 

The relation in Eq.\ (\ref{Suscept_Der}) is remarkable in the sense that it can replace all the partial derivatives with respect to the separate Hamiltonian matrix elements, $\{H_{ij}\}$, that are necessary to define the matrix elements of the susceptibility, $\{\chi_{ij}\}$, with a single forward directional derivative with respect to $\lambda$. 

Sometimes the perturbation $H^{(1)}$ and $A$ are of the same kind, for example, when we look at the polarizability, i.e. the changes in the dipole moment with respect to a perturbation from an external field that interacts with the system via the dipole operator. This makes it straightforward to calculate the different components of the polarizability tensor \cite{Weber04,VWeber05}. 

In cases where $A$ is a local observable and $H^{(1)}$ is a global perturbation we can take advantage of efficient sparse matrix algebra to calculate the susceptibility of $A$. Of course, the same property also holds for the opposite case, when $H^{(1)}$ is local and $A$ is a global observable. In this situation, the original recursive density matrix based perturbation scheme can also leverage sparse matrix algebra to accelerate the calculation of the density matrix perturbation \cite{ANiklasson04}.

\section{Generalizations} \label{Gen}

The susceptibility formulation presented above provides a fast dual approach to density matrix perturbation theory, which can be used to calculate linear response values of a given observable, as in Eq.\ (\ref{dual}), for a variety of time-independent perturbations in the Hamiltonian, $\{H^{(1)}_i\}$.
The theory was presented for integer occupations and assuming an orthonormal basis-set representation using the SP2 Fermi-operator expansion.

The density matrix based susceptibility relation in Eq.\ (\ref{Suscept_Der}), which can be calculated by the expansions in Eq.\ (\ref{DM2}) or  Eq.\ (\ref{DM3}), and the equivalence between these two expansions are some of the key results of this article. They rely either on the ``back propagation'' of the derivatives of an observable or on the matrix function relation in Eq.\ (\ref{MatFuncRel}). Our results are therefore quite general and should be applicable to a broad range of Fermi-operator expansion schemes beyond the recursive SP2 expansion presented here, including finite temperature expansions associated with canonical density matrix perturbation theory \cite{ANiklasson15} as well as to basis-set dependent perturbations for non-orthogonal representations. It is also possible to extend the dual susceptibility approach to self-consistent response calculations. These generalizations will be presented below.

\subsection{Non-orthonormal representations}

For non-orthonormal basis-set representations we have an overlap matrix, $S \ne I$, that we need to account for whenever we have basis-set-dependent perturbations, e.g.\ for the calculation of force terms or the Born-Effective charges. In these cases we need to distinguish between matrices in non-orthogonal atomic-orbital representations, $X$, and their orthonormalized matrix representations, $X^\perp$. Their relations are determined by the following congruence tranformations, 
\begin{align}
    &D = ZD^\perp Z^T,\\
    &A^\perp = Z^TAZ,\\
    &H^\perp = Z^THZ,
\end{align}
between the observable, $A$, Hamiltonian, $H$, and density matrix, $D$, in their atomic-orbital representation, and their corresponding orthonormal representations.
The orthonormal density matrix is then given by 
\begin{equation}
    D^\perp = \theta (\mu I - H^\perp),
\end{equation}
using the recursive density matrix expansion in Eq.\ (\ref{DM0}).

The transformation matrix, $Z$, above is given by an inverse factorization of the overlap matrix, where $Z$ satisfies the relation
\begin{equation}
    Z^TSZ = I.
\end{equation}
In the symmetric case, when $Z = Z^T$, we have that $Z = S^{-1/2}$, corresponding to the L\"{o}wdin orthonormalization. 

The expectation value, $a$, can be calculated either using the orthonormal or non-orthonormal atomic-orbital representations because of cyclic invariance of the trace operation, i.e.\
\begin{equation}
    a^{(0)} = {\rm Tr}\left[AD\right] \label{NonOrth_a_2}
    = {\rm Tr} \left[AZD^\perp Z^T\right] = {\rm Tr}\left[A^\perp D^\perp\right].
\end{equation}

In our presentation in previous sections we didn't need to make any distinction between the different matrix representation, i.e.\ between $X$ and $X^\perp$, because we assumed $S = I$, in which case $X = X^\perp$.

If we now calculate the derivative of the observable, $a$, with respect to atomic coordinates, $\{R_\tau \}$, we find that
\begin{align}
    &\frac{\partial a^{(0)}}{\partial R_\tau} = \frac{\partial {\rm Tr}\left[A ZD^\perp Z^T\right]}{\partial R_\tau} = {\rm Tr} \left[A_\tau D\right] +\\
    &+ {\rm Tr}\left[A\left(Z_\tau D^\perp Z^T + ZD^\perp_\tau Z^T + ZD^\perp Z_\tau ^T\right)\right], \label{PartDer}
\end{align}
where we use the notation $\partial Y \big / \partial R_\tau \equiv Y_\tau$. We can then use the relation (See Refs.\ \cite{ANiklasson07C,ANiklasson08b}),
\begin{equation}
Z_\tau = -(1/2)S^{-1}S_\tau Z,\label{Z_derivative}
\end{equation}
inserted into Eq.\ (\ref{PartDer}), which gives us
\begin{align}
    &\frac{\partial a^{(0)}}{\partial R_\tau} = {\rm Tr} \left[A_\tau D\right] + {\rm Tr} \left[A^\perp   D^\perp_\tau\right] + \nonumber \\
    &-\frac{1}{2}{\rm Tr}\left[AS^{-1}S_\tau D\right]
    -\frac{1}{2}{\rm Tr}\left[ADS_\tau S^{-1}\right]. \label{NonOrthDer}
\end{align}
In contrast to the previous orthonormal representation, we need to account for several position-dependent terms. Only the ${\rm Tr} \left[A^\perp   D^\perp_\tau\right]$ term stays the same for an orthonormal basis set representation, where $S = I$. However, all the other terms have basis-set dependent position derivatives that typically are straightforward to calculate in a similar manner to the terms that appears in the Pulay force calculation \cite{PPulay69,HSchlegel00,ANiklasson08b}. 

To calculate the remaining ${\rm Tr} \left[A^\perp   D^\perp_\tau\right]$ terms for perturbations over all the different atomic coordinates using density matrix perturbation theory could be quite expensive. For each perturbation we need to calculate a new density matrix response matrix. Additionally, in the self-consistent case, each density matrix response needs to be calculated using coupled perturbed SCF or density functional perturbation theory, which further increases the cost. With the proposed dual susceptibility formulation of density matrix perturbation theory we can instead use the susceptibility expression, where
\begin{equation}
    {\rm Tr} \left[A^\perp   D^\perp_\tau\right]  = {\rm Tr} \left[\chi^{A^\perp}   H_\tau^\perp \right].
\end{equation}
This dual approach may thus provide a significant simplification in determining the response for each new perturbation in the orthogonal Hamiltonian, which is given by
\begin{equation}
     H_\tau^\perp \equiv \frac{\partial (Z^THZ)}{\partial R_\tau} = Z_\tau^T H Z + Z^T H_\tau Z + Z^T H Z_\tau, 
\end{equation}
where we once again can use the relation in Eq.\ (\ref{Z_derivative}) for $Z_\tau$.

\subsection{Canonical susceptibility for fractional occupation numbers}

The relation in Eq.\ (\ref{MatFuncRel}) holds for any reasonable matrix function, $f(X)$, and is not limited to the step function with integer occupation numbers of the electronic states. In this way the density matrix response theory can be extended also to fractional occupation numbers at finite electronic temperatures, $T_e > 0$, where the Heaviside step function $\theta(\mu I - H)$ 
is replaced by the Fermi function,
\begin{equation}
    f(H) = \left(e^{\beta(H - \mu I)}+I \right)^{-1}.
\end{equation}
Here $\beta = 1/(k_{\rm B}T_e)$ is the inverse electronic temperature.
In this case we can use canonical density matrix perturbation theory for fractional occupation numbers \cite{ANiklasson15,ANiklasson20b} with the observable, $A$, as the perturbation to calculate the corresponding temperature-dependent susceptibility, i.e.\ as in Eq.\ (\ref{Suscept_Der}), where
\begin{equation}
  \chi^A(T_e) = \frac{d}{d \lambda} 
    \left(e^{\beta(H  - \mu I + \lambda (A+\mu_1 I))}+I \right)^{-1} \Big \vert_{\lambda = 0}.
\end{equation}
Also in this case there are fast recursive Fermi-operator expansion schemes \cite{ANiklasson15,SMMniszewski19,JFinkelstein21,JFinkelstein21B} that can take advantage of sparse matrix algebra or specialized AI-hardware. These schemes can also account for the linear response in the chemical potential, $\mu_1$ \cite{ANiklasson15,ANiklasson20b}. Alternatively, we may expand the Fermi function in a diagonal eigenbasis. We can then use the same recursive Fermi-operator scheme as above \cite{ANiklasson15,ANiklasson20b}, or we can use a modified Rayleigh-Schr\"{o}dinger perturbation theory adjusted for fractional occupation numbers, including the response in the chemical potential \cite{YNishimoto17}. However, the alternative formulations in a diagonal eigenbasis would require a diagonalization of the Hamiltonian, which is ill-suited for linear calculations using sparse matrix algebra or accelerated AI-hardware such as Tensor cores.

\subsection{Self-consistent susceptibility theory}

The relation in Eq.\ (\ref{MatFuncRel}) holds for a broad class of matrix functions, $f(X)$. This observation is also important if we would like to calculate the susceptibility functions within a self-consistent approach as in coupled perturbed SCF theory or density functional perturbation theory \cite{Gerratt68,JPople79,SBaroni01}. In this most general case the Fermi function, $f(X)$, is replaced by a recursive sequence of weighted Fermi functions that are mixed to find the self-consistent ground-state solution, both with respect to the density matrix and its response. In this generalized SCF case the relation in Eq.\ (\ref{MatFuncRel}) still holds. We can therefore calculate the self-consistent susceptibility in the same way as for the self-consistent density matrix response \cite{VWeber04,VWeber05}, but with the Hamiltonian perturbation replaced by the observable, $A$. In the appendix we give a schematic algorithm in Eq.\ (\ref{CP_SCF}) together with a derivation based on Hartree-Fock theory for coupled perturbed self-consistent susceptibility calculations. However, the methodology is quite general and is straightforward to apply also to other approximations beyond Hartree-Fock theory, including density functional perturbation theory. We will demonstrate the dual self-consistent susceptibility formulation of density matrix perturbation theory in the examples below using semi-empirical density functional theory.

\section{Examples}
The recursive susceptibility algorithm  as shown in Eq.\ (\ref{DMSuscept}) in the Appendix, which is based on Eq.\ (\ref{DM3}) merged with Eq.\ (\ref{DM0}) to avoid intermediate memory storage of $\{X_n\}$, provides a complementary dual approach to density matrix perturbation theory.  To illustrate the theory we first demonstrate its most general form, including self-consistent response calculation of the susceptibility with fractional occupation numbers, as shown in Eq.\ (\ref{CP_SCF}) in the Appendix.  We then adapt the algorithm in Eq.\ (\ref{DMSuscept}) to mixed precision floating point operations and perform test calculations using GPUs and Tensor cores to evaluate the performance. In our first examples we use self-consistent charge density functional tight-binding (SCC-DFTB) theory \cite{MElstner98,MFinnis98,TFrauenheim00,BAradi07,BHourahine20} for the self-consistent susceptibility calculations with fractional occupation numbers and then Hartree-Fock theory using a Gaussian basis set for the Tensor core demonstrations.

\begin{figure}
\includegraphics[scale=0.35]{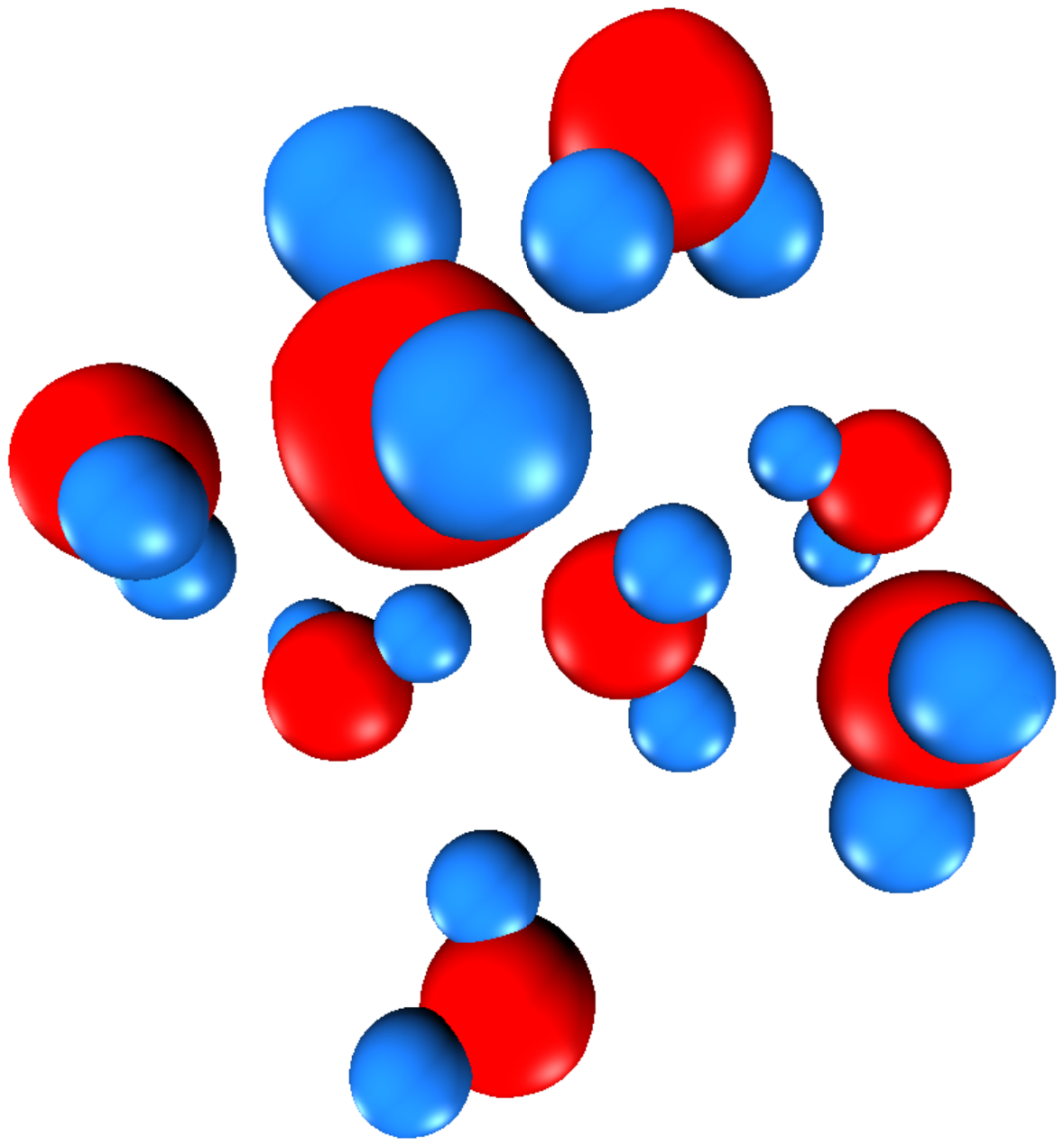}
\caption{\label{Fig_2}
{\small Water cluster with 8 molecules used to demonstrate the most general form of the susceptibility formulation of density matrix perturbation theory, including the self-consistent response and fractional occupation numbers. The numerical results using SCC-DFTB are shown in Table\ \ref{tab:SCF_dipoles}.}}
\end{figure}

\subsection{Self-consistent response calculations}

First we demonstrate the most general form of the dual susceptibility formulation of density matrix perturbation theory including the self-consistent response with fractional occupation numbers at elevated electronic temperatures and a non-orthogonal basis set representation. In our example we calculate the response in the dipole moment (in the $y$-direction) of a molecular system from a randomized local potential perturbation in the Hamiltonian. The calculations are performed in double precision both with the original density matrix perturbation theory and the dual susceptibility formulation to demonstrate their equivalence. The electronic structure calculations are based on self-consistent charge density functional tight-binding (SCC-DFTB) theory \cite{WHarrison80,MFoulkes89,DPorezag95,MElstner98,MFinnis98,TFrauenheim00,BAradi07,PKoskinen09,MGaus11,BAradi15,BHourahine20} as implemented in the LATTE electronic structure package \cite{LATTE,MCawkwell12,AKrishnapriyan17}. The self-consistent response calculations of the susceptibility at finite temperatures, $\chi^A$, can then be performed as in the Algorithm in Eq.\ (\ref{CP_SCF}) (see Appendix). In our comparison using the original density matrix perturbation theory we will use the Algorithm in Eq.\ (\ref{DM_CP_SCF}) (See Appendix), which corresponds to the approach of regular self-consistent density functional perturbation theory \cite{SBaroni01} or coupled perturbed self-consistent field calculations \cite{Gerratt68,JPople79,VWeber05}. In this case we use the recursive Fermi-operator expansion and the linear response calculation of the density matrix and susceptibility as given in Refs.\ \cite{ANiklasson15,ANiklasson20b}. 

Our test system is a cluster of 8 water molecules, see Fig.\ \ref{Fig_2}, and the perturbation, $H^{(1)}$, consists of local random shifts in the potential of each atom in the range of $[-1,1]$ eV. The calculations are run in double precision. We find that the calculated linear response in the dipole moment of the system using regular density matrix based coupled-perturbed self-consistent field or density functional perturbation theory \cite{Gerratt68,JPople79, SBaroni01,VWeber05} are equivalent up to 12 digits with the self-consistent dual susceptibility formulation of density matrix perturbation theory presented here. The results are shown in Table\ \ref{tab:SCF_dipoles} for three different electronic temperatures, $T_e = 300 $K, $T_e = 1,000 $K, and $T_e = 10,000 $K. Only the first 8 decimals are included. The two lower temperatures are low compared to the electronic (HOMO-LUMO) gap of the water cluster and the dipole moment of the water cluster does not change at $T_e < 1000 K$. Only at 10,000 K do we find a significant shift in the dipole moment because of fractional occupation numbers. The results clearly demonstrate the numerical equivalence between the dual approaches to density matrix based linear response theory.

\begin{table}[]
    \centering
    \begin{tabular}{l|c|c|c}
      \hline
         {\rm Temperature} & 300 K & 1,000 K & 10,000 K \\
           \hline
         ${\rm Tr}[AD^{(1)}]$ &  ~-0.15820523~ & ~-0.15820523 ~& ~-0.16488875 \\
        ${\rm Tr}[\chi^AH^{(1)}]$~ &  ~-0.15820523 ~& ~-0.15820523 ~& ~-0.16488875 

    \end{tabular}
    \caption{The self-consistent linear response in the total dipole moment in an arbitrarily chosen direction ($y$-direction) of a small water cluster  (see Fig.\ \ref{Fig_2}) calculated with SCC-DFTB theory in double precision.  The dipole moments are in units of Debye with respect to a random perturbation in the Hamiltonian (in eV), where the potential of each atom is shifted randomly (with a uniform distribution) in the range of $[-1,1]$ eV. The self-consistent linear response calculation of the susceptibility matrix $\chi^A$ was calculated as in the Algorithm in Eq.\ (\ref{CP_SCF}) at three different electronic temperatures. The corresponding self-consistent linear response calculation of $D^{(1)}$ was based on density matrix perturbation theory as in the Algorithm in Eq.\ (\ref{DM_CP_SCF}) \cite{Gerratt68,JPople79,VWeber05}. The tightest possible level of SCF convergence with double precision arithmetic was used. We find that the calculated dipole response value of the two approaches are the same for the first 12 digits (only 8 digits are shown), which demonstrates the numerical equivalence between the complementary dual formulations.}
    \label{tab:SCF_dipoles}
\end{table}

\subsection{Mixed precision matrix multiplications}

The computational kernel of Eq.\ (\ref{DM1}) and Eq.\ (\ref{DM3}) that dominates the cost in the dual response calculations is a generalized dense matrix-matrix multiplication. These can be performed with high-efficiency using AI-hardware that is optimized for the tensor contractions in convolutional deep neural network calculations. In general these tensor contractions are performed using low precision floating point multiplies and higher precision accumulations -- for example, half precision representations for the input with single precision accumulation of the results. We can adapt this mixed precision format for density matrix susceptibility calculations, where we represent a matrix $X$ given in single precision as,
\begin{equation}
    X = X^{(h)} + X^{(l)}, \label{doubleHalf}
\end{equation}
in near single precision by two half precision floating point matrices. Here 
\begin{align}
    X^{(h)} &= {\rm FP16}[X]\;,\\
    X^{(l)} &= {\rm FP16}[X-{\rm FP32}[X^{(h)}]]\;,
\end{align}
are the matrices containing the higher ($h$) and lower ($l$) significant parts of the floating point representation, and where ${\rm FP16}[\cdot]$ denotes the half precision representation and ${\rm FP32}[\cdot]$ denotes single precision representation. A matrix product can then be performed by four separate half precision multiplications,
\begin{align}
   X &\times Y = \left(X^{(h)} + X^{(l)}\right)\left( Y^{(h)} + Y^{(l)}\right)\\
    &= X^{(h)}Y^{(h)} + X^{(h)}Y^{(l)} + X^{(l)}Y^{(h)} + X^{(l)}Y^{(l)}.\label{MixPrecMult}
\end{align}
If $X = Y$ and $X = X^T$ we can use symmetry to avoid one of the multiplications. Additionally, in the particular case of multiplication with FP16 data and dot product accumulation in FP32, formal component-wise error bounds suggest that the $X^{(l)}Y^{(l)}$ can be discarded without affecting accuracy \cite{MFasi23}. Thus the multiplication in Eq.~(\ref{MixPrecMult}) only requires two matrix multiplications instead of four. In the more general case, where $X \ne Y$, we need three separate multiplications. Thus to calculate $D_0$ and $D_1$ using Eq.~(\ref{DM0}) and Eq.~(\ref{DM1}), we need a total of 5 multiplications at each iteration to compute the necessary matrix multiplications since,
\begin{align}\label{eq:mults}
    \begin{split}
    X_n^2 &\approx {X_n}^{(h)}{X_n}^{(h)} +{X_n}^{(h)}{X_n}^{(l)} \\
     & \hspace{3cm}+({X_n}^{(h)}{X_n}^{(l)})^T\\
    Y_n X_n & \approx {Y_n}^{(h)}{X_n}^{(h)} +{Y_n}^{(h)}{X_n}^{(l)} \\
     & \hspace{3cm}+{X_n}^{(h)}{Y_n}^{(l)} \;,
    \end{split}
\end{align}
and $(Y_nX_n)^T=X_nY_n$ by symmetry. Both the ${Y_n}^{(l)}{X_n}^{(l)}$ and ${X_n}^{(l)}{X_n}^{(l)}$ terms are discarded. 

\subsection{Susceptibility calculations using GPUs and Tensor cores}

\begin{figure}
\includegraphics[scale=0.25]{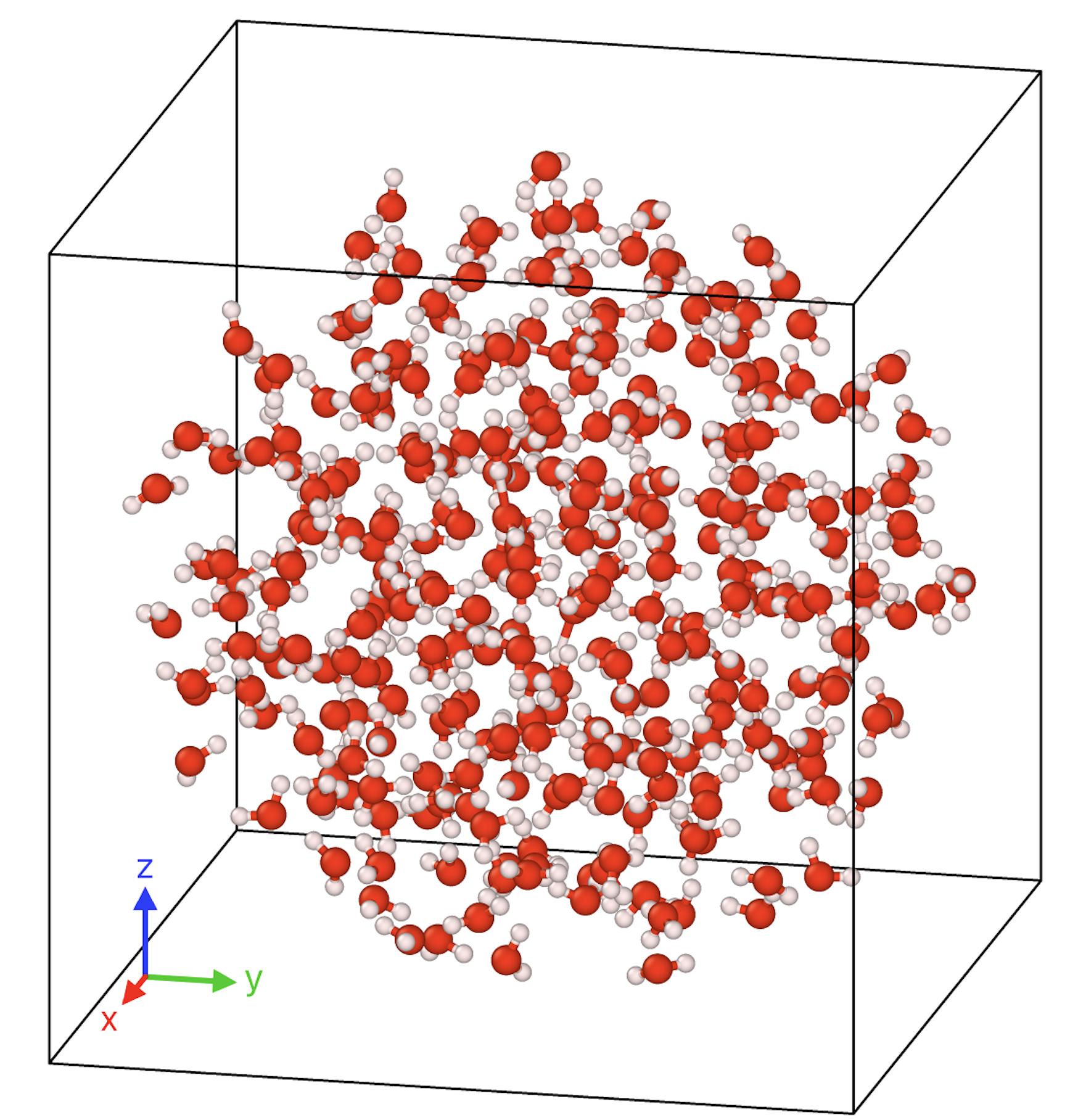}
\caption{\label{water903Atoms}
{\small A cluster of 301 water molecules where each atom has 8 basis functions, corresponding to the $N=7,224$ test case in Table~\ref{tab:TC_susceptibility}.}}
\end{figure}

To test and evaluate the performance of the density matrix susceptibility calculations with GPUs and Tensor cores we use Eq.\ (\ref{DM3}) merged with Eq.\ (\ref{DM0}) as given by the Algorithm in Eq.\  (\ref{DMSuscept}) (See Appendix). For the GPU calculations we use single-precision arithmetics and for the Tensor cores we use the double half precision representation of the matrices as in Eq.\ (\ref{doubleHalf}), with the mixed precision approach to the dense matrix-matrix multiplications as in Eq.\ (\ref{MixPrecMult}). 
The simulations were performed for water clusters of varying sizes, with 100, 190 and 301 molecules where each atom has 8 basis functions, making $N$ = number of atoms $\times$ number of basis functions. For example, in Fig.~\ref{water903Atoms} we show the 301 water-molecule cluster that corresponds to $N = 7,224$ for the dimension of our matrices in the test calculations. We calculate the local response in the dipole moment ($x$-direction) that is driven by a random perturbation in the Hamiltonian, i.e. in the same way as in our previous examples in Table\ \ref{tab:SCF_dipoles}. The electronic structure is approximated by restricted Hartree-Fock theory using the Gaussian 6-31G$^{**}$ basis set. Fockian matrices (i.e.\ the effective single-particle Hamiltonians) were generated (self-consistently) for the water clusters with the electronic structure software package Ergo \cite{Ergo}. The water clusters were extracted from a molecular dynamics simulation of bulk water at ambient temperature and pressure by including all water molecules within spheres of varying radii. The water cluster geometries are available at \url{ergoscf.org}. Our expectation values are the net dipole moment (in Debye) of the water clusters determined by the 
dipole moment operator that couples the system to an external field in an arbitrarily chosen direction ($x$-direction).  As a perturbation we use a randomized local potential perturbation in the same way as in our previous example using SCC-DFTB theory. Only the Fockian is calculated self-consistently, not the linear response in the density matrix nor the susceptibility in the dual approach.

Calculations were run on the Venado supercomputer at Los Alamos National Laboratory, which is based on the Nvidia Grace-Hopper superchip architecture. A single superchip combines one Nvidia Grace CPU with an H100 GPU using a coherent shared-memory architecture. Each H100 GPU contains 4th generation Tensor cores \cite{nvda-h100} and has a peak theoretical deep learning performance (FP16/BFLOAT16) of 495 Tflops. Counting of flops is done by combining additions and multiplications as a single floating-point operation. 

Numerical results are shown in Table\ \ref{tab:TC_susceptibility} using both mixed precision with Tensor cores and pure single precision (FP32) with GPUs using standard Nvidia CUDA compute cores. The Nvidia cuBlas library was used for all matrix multiplications and Tensor cores were accessed through cuBlas using the {\tt GemmEx} function. The dipole moments are measured in Debye with the randomized Hamiltonian perturbations in eV. The first two lines of the table show the linear response in the dipole moment calculated on Tensor cores (TC) using the mixed precision approach, either with density matrix perturbation theory, ${\rm Tr}[D^{(1)}A]$, or with the dual susceptibility approach, ${\rm Tr}[\chi^AH^{(1)}]$. The next two lines contain the corresponding GPU calculations using pure single precision arithmetic as a reference. We find some interesting behavior. In particular, the regular density matrix perturbation approach gives poor results because of the low numerical precision. The mixed precision approach running on the Tensor cores is simply not accurate enough to capture the correct result. However, the dual susceptibility approach shows a much higher accuracy, which is within 5\% of the FP32 reference data. Moreover, these susceptibility based response calculations can be performed with a peak performance of over 200 Tflops using the Tensor cores on a single H100 GPU. This is an order of magnitude higher than the GPU FP32 reference calculations. 

If the Tensor core results are not sufficiently accurate they can be enhanced in accuracy using additional refinement \cite{JFinkelstein21B} or they can be used as an accurate initial guess to higher-order precision calculations, for example, in the last iterations of a self-consistent optimization \cite{Shang21}.  We cannot assume that the susceptibility formulation will always provide the more accurate result over density matrix perturbation theory. However, the examples in Table\ \ref{tab:TC_susceptibility}, demonstrate how the dual susceptibility formulation sometimes may complement the density matrix perturbation calculations if they fail. Another interesting possibility may be to utilize alternative mixed precision representations to the straightforward splitting scheme above, for example, the Ozaki scheme as in \cite{WDawson24,KOzaki12}. With this technique it is possible to adjust the numerical accuracy to the needed precision, though it may require a significant increase in the number of matrix multiplications.

In our simulations we achieve a peak performance of about 220 Tflops using the Tensor cores, which is less than 50\% of the theoretical peak. To get higher performance we would need larger system sizes \cite{JFinkelstein21,JFinkelstein21B,JFinkelstein22}. In our previous work using A100 Tensor cores, the performance, which is governed by the dense matrix multiplications, doubled when going from matrix sizes of around $7,000 \times 7,000$ to $20,000 \times 20,000$ \cite{JFinkelstein21}.  However, for sufficiently large system sizes, linear scaling methods using sparse matrix algebra may become more efficient. We have also noticed fluctuations in the performance between different compute nodes on the newly built Venado supercomputer.  Table\ \ref{tab:TC_susceptibility} therefore only shows the best performance measurements over several test calculations. 

\begin{table}[]
    \centering
    \begin{tabular}{l|c|c|c}
    \hline
         {\rm Matrix Size ($N \times N$)} & $N=2400$ & $N=4560$ & $N=7224$ \\
         \hline
        {\rm TC} ${\rm Tr}[D^{(1)}A]$ &  6.1132 & 3.3647 & 6.5898  \\
        {\rm TC} ${\rm Tr}[\chi^AH^{(1)}]$ & 2.3589  & 3.2138  & 2.5200  \\
        {\rm GPU} {\rm FP32} ${\rm Tr}[D^{(1)}A]$ &  2.2481 & 3.1226 & 2.6899 \\
        
        {\rm GPU} {\rm FP32} ${\rm Tr}[\chi^AH^{(1)}]$ & 2.2440  & 3.1222  & 2.6919\\
        Peak TC (Tflops) & 120.8 & 203.4 & 222.4 \\
        Peak GPU (Tflops) & 18.6 & 22.9 & 22.9 \\
    \end{tabular}
    \caption{First order response in the dipole moments (in the $x$-direction) for water clusters of various sizes with respect to a randomized Hamiltonian perturbation using restricted Hartree-Fock theory with a Gaussian 6-31G** basis set. Values in the dipole response are calculated either with density matrix perturbation theory, ${\rm Tr}[D^{(1)}A]$, or with the dual susceptibility approach, ${\rm Tr}[\chi^AH^{(1)}]$. Peak 
    performance over several tests in Tflops (accounting for fused multiply-adds) are displayed for ${\rm Tr}[\chi^AH^{(1)}]$ on the bottom line. A schematic picture of the water cluster in the $N = 7,224$ test case is shown in Fig.~\ref{water903Atoms}.}
\label{tab:TC_susceptibility}
\end{table}

\section{Summary and discussion}

We have introduced a dual susceptibility approach to density matrix perturbation theory, where the linear response in an expectation value is calculated using the susceptibility instead of the density matrix response, as is schematically illustrated in Fig.\ \ref{Fig_1}. The calculation of the susceptibility can be performed with the same recursive expansion schemes as in density matrix perturbation theory, but where observable takes the place of a perturbation. This can be achieved, either in a backward or in a forward expansion. The dual susceptibility formulation provides an alternative that complements previous density matrix perturbation theory. The dual susceptibility approach is particularly useful when multiple response calculations are needed for the same observable under different Hamiltonian perturbations. The dual approach would also be useful for response calculations of an observable corresponding to local (single atom) response arising from a non-local Hamiltonian perturbation. For example, the change in the local net charge of a specific atom due do a uniform external field. In this case sparse matrix algebra can be used to accelerate the susceptibility calculation.

The recursive susceptibility expansion algorithms, Eqs.\ (\ref{DM2}) and (\ref{DM3}), and their equivalence, as well as the generalization to fractional occupation numbers and self-consistent linear response theory, are the key results of this article. The SP2 Fermi-operator expansion was used to illustrate the susceptibility formulation of density matrix perturbation theory, but the theory is general and applies to a broad range of recursive Fermi-operator expansion methods. This also includes finite temperature expansions with fractional occupation numbers (see Appendix), diagonal eigenbasis representations, and self-consistent susceptibility calculations as in coupled perturbed self-consistent field or density functional perturbation theory. 
This was also validated in our test examples that demonstrated the numerical equivalence between the original recursive density matrix perturbation theory and its susceptibility formulation for the most general case, including self-consistency and fractional occupation numbers at finite electronic temperatures.

The dual susceptibility formulation could also be applicable as a complement to other forms of density matrix perturbation theory, for example, where the linear response density matrices appear as solutions to Sylvester-like commutator (or super-operator) equations \cite{COchsenfeld04,Liang05,JKussmann07,Ringholm14,Kussman15} or as a complement to regular density functional perturbation theory and coupled-perturbed self-consistent field calculations.

The susceptibility formulation of linear response theory, based on recursive density matrix perturbation theory, enables fast, repeated AI-hardware or GPU-based calculations of the response in the expectation values of an observable under different perturbations to the Hamiltonian. We believe this could be highly useful, for example, in the optimization of semi-empirical Hamiltonians in self-consistent electronic structure theory using machine-learning methods \cite{DYaron18,PDral20,ZGuoqing22,DYaron23}.  

The susceptibility formulation of density matrix perturbation theory is based on repeated generalized matrix-matrix multiplications. These operations can be performed using numerically thresholded sparse matrix algebra to achieve linear scaling complexity for sufficiently large sparse systems. The recursive expansions are also well adapted for dense matrix algebra that can take advantage of GPUs and modern AI-hardware that are optimized for tensor contractions. However, the accuracy of the calculations may be limited. In our examples we demonstrated a performance of over 20 TFlops for the GPU FP32 calculations and over 220 Tflops using the Tensor cores of an Nvidia H100 GPU for the recursive susceptibility calculations.

\section{Acknowledgements}
This work is supported by the U.S. Department of Energy Office of Basic Energy Sciences (FWP LANLE8AN), the LANL LDRD-DR program, by the U.S. Department of Energy through the Los Alamos National Laboratory, and by the Swedish national strategic e-science research program (eSSENCE). We thank the CCS-7 group and the Darwin cluster at Los Alamos National Laboratory for computational resources. Darwin is funded by the Computational Systems and Software Environments (CSSE) subprogram of LANL’s ASC program (NNSA/DOE). Los Alamos National Laboratory is operated by Triad National Security, LLC, for the National Nuclear Security
Administration of the U.S. Department of Energy Contract No. 892333218NCA000001.

\section{Author Declarations}

\subsection{Conflict of interest}

The authors have no conflicts to disclose.

\section{Appendix}

In the Appendix we summarize some of the algorithms, including the recursive susceptibility calculations and the corresponding density matrix response algorithms. We also present the generalized algorithms for the self-consistent calculations including fractional occupation numbers at finite electronic temperatures. The thermal Hartree-Fock formalism is used in the algorithms for the self-consistent calculations, which is easily generalized to Kohn-Sham density functional perturbation theory or various approximate semi-empirical formulations. Thereafter, in some detail, we show the motivations behind the dual susceptibility relations used in the derivations of the algorithms.

In the algorithms and discussions below we assume a symmetric matrix representation of the observable, i.e., $A = A^T$, if not $A$ and $A^T$ are included separately. 

\subsection{Algorithms}

\subsubsection{Recursive susceptibility calculation}

Merging the recursive SP2 Fermi-operator expansion in Eq.\ (\ref{DM3}) with Eq.\ (\ref{DM0}) to avoid memory storage of the density matrix expansion, $\{X_n\}$, the susceptibility algorithm (assuming $A = A^T$) takes the form 
  \begin{align} \label{DMSuscept}
     & X_1 = \alpha I + \beta H^{(0)} \nonumber \\
     &Y_1 = \beta A \nonumber\\
     &{\rm for~} n = 1 {\rm ~to~} M\nonumber \\
     & ~~~~ Y_{n+1} = (1-\sigma_n)Y_n + \sigma_n(Y_nX_n + X_nY_n)\\
     & ~~~~ X_{n+1} = (1-\sigma_n)X_n + \sigma_nX_n^2 \nonumber\\
     &{\rm end}\nonumber \\
     &D^{(0)} = X_{M+1} \nonumber \\
     &\chi^A = Y_{M+1} \nonumber\\
     &a^{(0)} = {\rm Tr} \left[ D^{(0)} A \right] \nonumber \\
     &a^{(1)} = {\rm Tr} \left[\chi^A H^{(1)}  \right]. \nonumber
 \end{align}
How the sequence of $\{\sigma_n = \pm 1\}$ is chosen and how convergence is determined is given by Alg.\ 1 in Ref.\ \cite{JFinkelstein21}. 

\subsubsection{Density matrix response calculation}

The density matrix susceptibility algorithm above in Eq.\ (\ref{DMSuscept}) is closely aligned with the corresponding density matrix perturbation scheme,
  \begin{align} \label{DMPert}
     & X_1 = \alpha I + \beta H^{(0)} \nonumber \\
     &Y_1 = \beta H^{(1)} \nonumber\\
     &{\rm for~} n = 1 {\rm ~to~} M\nonumber \\
     & ~~~~ Y_{n+1} = (1-\sigma_n)Y_n + \sigma_n(Y_nX_n + X_nY_n)\\
     & ~~~~ X_{n+1} = (1-\sigma_n)X_n + \sigma_nX_n^2 \nonumber\\
     &{\rm end}\nonumber \\
     &D^{(0)} = X_{M+1} \nonumber \\
     &D^{(1)} = Y_{M+1} \nonumber \\
     &a^{(0)} = {\rm Tr} \left[D^{(0)}A\right] \nonumber \\ 
     &a^{(1)} = {\rm Tr} \left[D^{(1)}A\right].\nonumber
 \end{align}
 
How the sequence of $\{\sigma_n = \pm 1\}$ is chosen and how convergence is determined is given by Alg.\ 1 in Ref.\ \cite{JFinkelstein21}.

\subsubsection{Self-consistent susceptibility}

Our susceptibility formulation of density matrix perturbation theory can easily be adapted to self-consistent calculations. Here we will give an explicit example for restricted thermal Hartree-Fock theory, where $G(D) = 2J(D) - K(D)$ is the two-electron Coulomb and exchange term  \cite{ASzabo89,NMermin63,ANiklasson21b}.
In this case the self-consistent susceptibility, $\chi$, for the observable, $A$, using a non-orthogonal basis-set representation with overlap matrix, $S$, with the inverse factorization, $Z^TSZ = I$, is given by the coupled perturbed self-consistent field algorithm,
\begin{align} \label{CP_SCF}
    &\chi = 0 \nonumber\\
    &{\rm while ~}\|\Delta \chi \|~ > ~ \epsilon \nonumber \\
    &~~~\chi_{\rm old} = \chi \nonumber \\
    &~~~ A_1^\perp = Z^T\left(A + G(\chi)\right)Z \nonumber \\
    &~~~ \chi^\perp = \frac{d}{d \lambda} \left(  
    e^{\left[\beta(H_0^\perp - \mu_0 I + \lambda (A_1^\perp - \mu_1 I)) \right]}+I
    \right)^{-1} \Big \vert_{\lambda = 0}\\
    &~~~ \chi = Z \chi^\perp Z^T \nonumber\\
    &~~~ \Delta \chi  = \chi - \chi_{\rm old} \nonumber\\
    &~~~ \chi = \chi_{\rm old} + c_{\rm mix} \Delta \chi\nonumber \\
   &{\rm end} \nonumber \\
   &\chi^A = \chi \nonumber\\
   &a^{(1)} = {\rm Tr} \left[\chi^A H^{(1)}  \right]. \nonumber
\end{align}
Here the quantum-response calculation is given for a general Fermi function with fractional occupations numbers, where the inverse temperature, $\beta = 1/(k_{\rm B}T)$. The derivative with respect to $\lambda$ of the Fermi function that determines $\chi^\perp$ in each iteration can be calculated recursively as in Ref.\ \cite{ANiklasson15} or as in Alg.\ 2 in Ref.\ \cite{ANiklasson20b}, including the important linear response in the chemical potential, $\mu_1$, i.e.\ where $\mu = \mu_0 + \lambda \mu_1$. We then chose $\mu_1$ such that ${\rm Tr}[\chi] = 0$. The calculations have converged when the residual, $\Delta \chi$, is less than some chosen tolerance, $\epsilon$. At zero electronic temperature (as $\beta \rightarrow \infty$) the algorithm in Eq.\ (\ref{DMSuscept}) can be used instead to calculate $\chi^\perp$ in each iteration. In this case we have integer occupation with a finite electronic gap and $\mu_1 = 0$. To achieve SCF convergence we use a simple linear mixing scheme with a constant mixing parameter, $c_{\rm min}$. More advanced SCF acceleration techniques can be used alternatively \cite{PPulay80,PPulay82,VWeber03}. At convergence the self-consistent linear response in the expectation values, $a^{(1)}$, are easily obtained for a wide range of perturbations in the Hamiltonian $H^{(1)}$. 

\subsubsection{Self-consistent density matrix response}

The self-consistent  susceptibility algorithm above in Eq.\ (\ref{CP_SCF}) can be compared to the corresponding self-consistent density matrix perturbation scheme for a first-order perturbation, $H^{(1)}$, in the Hamiltonian,
\begin{align} \label{DM_CP_SCF}
    &D_1 = 0 \nonumber\\
    &{\rm while ~}\|\Delta D_1 \|~ > ~ \epsilon \nonumber \\
    &~~~D_1^{\rm old} = D_1 \nonumber \\
    &~~~ H_1^\perp = Z^T\left(H^{(1)} + G(D_1)\right)Z \nonumber \\
    &~~~ D_1^\perp = \frac{\partial}{\partial \lambda} \left(  
    e^{\left[\beta(H_0^\perp - \mu_0 I + \lambda (H_1^\perp - \mu_1 I)) \right]}+I
    \right)^{-1} \Big \vert_{\lambda = 0}\\
    &~~~ D_1 = Z D_1^\perp Z^T \nonumber\\
    &~~~ \Delta D_1  = D_1 - D_1^{\rm old} \nonumber\\
    &~~~ D_1 = D_1^{\rm old} + c_{\rm mix} \Delta D_1 \nonumber \\
   &{\rm end} \nonumber \\
   & D^{(1)} = D_1 \nonumber \\
   &a^{(1)} = {\rm Tr} \left[D^{(1)} A  \right] \nonumber.
\end{align}
Also here the notation is based on restricted thermal Hartree-Fock theory, where $G(D) = 2J(D) - K(D)$, which includes the two-electron Coulomb and exchange integrals \cite{ASzabo89}.
 The derivative $D_1^\perp$ with respect to $\lambda$ of the Fermi function that is calculated in each iteration can be calculated recursively as in Ref.\ \cite{ANiklasson15} or as in Alg.\ 2 in Ref.\ \cite{ANiklasson20b}, including the important linear response in the chemical potential, $\mu_1$, i.e.\ where $\mu = \mu_0 + \lambda \mu_1$.
At convergence the self-consistent linear response in the density matrix, $D_1$, can be used to calculate the response for various observables.
In Hartree-Fock theory this self-consistent scheme is often called the coupled perturbed self-consistent field approach \cite{Gerratt68,JPople79,VWeber05}, or alternatively, density functional perturbation theory \cite{SBaroni01}, when based on Kohn-Sham density functional theory.  For the SCF convergence we here used the same simple linear mixing scheme with a constant mixing parameter, $c_{\rm mix}$, as above in Eq.\ (\ref{CP_SCF}). More advanced SCF acceleration techniques can be used also in this case, e.g.\ see Refs.\ \cite{PPulay80,PPulay82,VWeber03}.

\subsection{Some important relations behind the dual susceptibility formulation}

\subsubsection{ Eq.\ (\ref{MatFuncRel}), non-self-consistency }

Here we will show the relation in Eq.\ (\ref{MatFuncRel}), i.e.\
 \begin{equation} \label{FuncRel}
\frac{\partial {\rm Tr}\left[Af(X)\right]  }{\partial X} =  
\frac{d f(X + \lambda A^T)}{d \lambda}\Big \vert_{\lambda = 0},
 \end{equation}
holds, assuming $X = X^T$,  for any general matrix polynomial function, $f(X) = \sum_{n=0}^M c_n X^n$, and then we give a proof of the closely related relation in Eq.\ (\ref{Equivalence}). 

It is first easy to show that Eq.\ (\ref{FuncRel}) holds for any first-order matrix polynomial, $f(X) = c_0I + c_1X$, by direct insertion. For higher-order polynomial terms, where $f(X) = X^n$ with $n \ge 2$, we then have that the right-hand side of Eq.\ (\ref{FuncRel}) is 
\begin{align}
    f(X+& \lambda A^T) = 
    (X+\lambda A^T)^n \nonumber \\
    &= X^n + \lambda \left(X^{n-1}A^T+X^{n-2}A^T X +\right.\\ 
    &\left. + X^{n-3}A^TX^2+ \ldots +A^TX^{n-1}\right) + {\cal O}(\lambda^2).\nonumber
\end{align}
This means that
\begin{align}
\frac{\partial f(X + \lambda A^T)}{\partial \lambda}&\Big \vert_{\lambda = 0} = X^{n-1}A^T+ X^{n-2}A^T X +\\
&+  X^{n-3}A^T X^2 +\ldots + A^TX^{n-1},\nonumber
\end{align}
which is easy to identify with the left-hand side of Eq.\ (\ref{FuncRel}) with $f(X) = X^n$, where 
\begin{align}
    &\frac{\partial {\rm Tr} \left[ AX^n\right]}{\partial X_{ij}}  =  
    X^{n-1}A^T+  \ldots + A^TX^{n-1}
\end{align}
if we assume $X=X^T$. In this way it is easy to see how Eq.\ (\ref{FuncRel}) holds for any function, $f(X)$, that can be described by a matrix polynomial.

\subsubsection{ Eq.\ (\ref{Equivalence}), self-consistency}

We will show here that our susceptibility formulation holds, i.e.\ equality between the expressions in Eq.~\eqref{direct} and Eq.~\eqref{dual}, where
\begin{align}
    {\rm Tr} \left[D^{(1)}A\right] &= {\rm Tr} \left[\chi^A H^{(1)}\right].
\end{align}
We consider the general setting with {\it self-consistent} response and a Fermi occupation function, $f$, with sufficient regularity on the spectrum of the unperturbed Hamiltonian, $H^{(0)}$, where
\begin{align}
D^{(1)} & = \frac{d}{d\lambda} \left.f(H^{(0)}+\lambda (H^{(1)}+G(D^{(1)})))\right|_{\lambda=0} 
\end{align}
and 
\begin{align}
\chi^A & = \frac{d}{d\lambda} \left.f(H^{(0)}+\lambda (A+G(\chi^A)))\right|_{\lambda=0}. 
\end{align}
We then have that 
\begin{align}
    {\rm Tr} \left[D^{(1)}A\right] =& {\rm Tr} \left[D^{(1)}(A+G(\chi^A))\right]  \\
    & -{\rm Tr} \left[D^{(1)}G(\chi^A)\right] \nonumber
\end{align}
and
\begin{align}
    {\rm Tr} \left[\chi^A H^{(1)}\right] = & {\rm Tr} \left[\chi^A(H^{(1)}+G(D^{(1)}))\right] \\
    & -{\rm Tr} \left[\chi^A G(D^{(1)})\right]\nonumber
\end{align}
and the desired equality therefore holds if 
\begin{align}\label{eq:susceptibility_proof_part_1}
{\rm Tr} \left[D^{(1)}(A+G(\chi^A))\right] = 
{\rm Tr} \left[\chi^A(H^{(1)}+G(D^{(1)}))\right]
\end{align}
and 
\begin{align}\label{eq:susceptibility_proof_part_2}
{\rm Tr} \left[D^{(1)}G(\chi^A)\right] =
{\rm Tr} \left[\chi^A G(D^{(1)})\right].
\end{align}

We will first show that Eq.\ \eqref{eq:susceptibility_proof_part_1} holds.
Let,  $H^{(0)} = V\Lambda V^*$ with diagonal $\Lambda =\mathrm{diag}(\lambda_1,\dots,\lambda_N)$ and let $L$ be the Loewner matrix with entries given by the divided differences $L_{i,j} = f[\lambda_i,\lambda_j]$. Then, by Dalecki{\u\i} and Kre{\u\i}n~\cite{DaleckiiKrein_1965},
\begin{align}
D^{(1)} & = V(L\circ (V^*(H^{(1)}+G(D^{(1)}))V))V^*, \\
\chi^A  & = V(L\circ (V^*(A+G(\chi^A))V))V^*, 
\end{align}
where $\circ$ is the Hadamard or elementwise product.
Since the trace of matrix products is invariant to cyclic permutation we have that
\begin{align} 
    & {\rm Tr} \left[D^{(1)}(A+G(\chi^A))\right] = \\
    & {\rm Tr} \left[(L\circ (V^*(H^{(1)}+G(D^{(1)}))V))V^*(A+G(\chi^A))V\right]
    \nonumber
\end{align}
and 
\begin{align}
& {\rm Tr} \left[\chi^A(H^{(1)}+G(D^{(1)}))\right] = \\
    & {\rm Tr} \left[ 
    (L\circ (V^*(A+G(\chi^A))V))V^*(H^{(1)}+G(D^{(1)}))V\right].
    \nonumber
\end{align}
These two expressions are equal since 
\begin{align}
{\rm Tr} \left[(L \circ X) Y\right] = {\rm Tr} \left[(L \circ Y) X\right]
\end{align}
holds with symmetric $X$ and $Y$.

In the case of Hartree-Fock theory $G(D^{(1)})$ is the contraction with the two-electron integrals and Eq.\ \eqref{eq:susceptibility_proof_part_2} holds. In a more general case we assume that $G(D^{(1)})$ is symmetric  and depends linearly on $D^{(1)}$. In the most general case we may use an expression 
$G(D^{(1)}) = AD^{(1)} + D^{(1)}A + \sum_i (B_i D^{(1)} C_i + C_i D^{(1)} B_i)$ for some set of matrices (or supermatrices) $A$, $\{B_i\}$, and $\{C_i\}$. Because the trace is invariant under cyclic permutations, Eq.\ \eqref{eq:susceptibility_proof_part_2} holds also in this case as well as Eq.\ \eqref{eq:susceptibility_proof_part_1}.

\bibliography{DMSusceptibility}

\end{document}